\documentclass[11pt]{article}

\usepackage{amssymb,amsmath,amsthm,amsfonts}
\usepackage[numbers,comma,sort&compress]{natbib}
\usepackage[margin=1in]{geometry}
\usepackage[colorlinks]{hyperref}
\usepackage[all]{hypcap}
\usepackage{url}
\usepackage{authblk}
\usepackage[percent]{overpic}
\usepackage[font=small]{caption}
\usepackage[labelformat=simple]{subcaption}

\input{Qcircuit}
\let\bra\relax
\let\ket\relax

\usepackage{mathtools}
\mathtoolsset{centercolon}
\DeclarePairedDelimiter\bra{\langle}{\rvert}
\DeclarePairedDelimiter\ket{\lvert}{\rangle}
\DeclarePairedDelimiterX\braket[2]{\langle}{\rangle}{#1 \delimsize\vert #2}

\newcommand{\eq}[1]{(\ref{eq:#1})}
\renewcommand{\sec}[1]{\hyperref[sec:#1]{Section~\ref*{sec:#1}}}
\newcommand{\fig}[1]{\hyperref[fig:#1]{Figure~\ref*{fig:#1}}}
\newcommand{\subfig}[1]{\protect\subref{fig:#1}}

\renewcommand{\d}{\mathrm{d}}
\newcommand{\R}{\mathbb{R}}

\DeclareMathOperator{\sgn}{sgn}
\DeclareMathOperator{\spn}{span}
\DeclareMathOperator{\MF}{MF}

\newcommand{\nlin}{\kappa}
\newcommand{\nlinred}{\bar\nlin}
\newcommand{\natoms}{\mathcal{N}}

\begin{document}

\title{Optimal state discrimination and unstructured \\ search in nonlinear quantum mechanics}

\author{Andrew M.\ Childs${}^{1,2,3}$ and Joshua Young${}^{3,4}$ \\
\small{$^{1}$Department of Computer Science, Institute for Advanced Computer Studies, and \\ Joint Center for Quantum Information and Computer Science, University of Maryland} \smallskip\\
\small{$^{2}$Department of Combinatorics \& Optimization, University of Waterloo} \smallskip\\
\small{$^{3}$Institute for Quantum Computing, University of Waterloo} \smallskip\\
\small{$^{4}$Cheriton School of Computer Science, University of Waterloo}}

\date{}

\maketitle


\begin{abstract}
Nonlinear variants of quantum mechanics can solve tasks that are impossible in standard quantum theory, such as perfectly distinguishing nonorthogonal states.  Here we derive the optimal protocol for distinguishing two states of a qubit using the Gross-Pitaevskii equation, a model of nonlinear quantum mechanics that arises as an effective description of Bose-Einstein condensates.  Using this protocol, we present an algorithm for unstructured search in the Gross-Pitaevskii model, obtaining an exponential improvement over a previous algorithm of Meyer and Wong.  This result establishes a limitation on the effectiveness of the Gross-Pitaevskii approximation.  More generally, we demonstrate similar behavior under a family of related nonlinearities, giving evidence that the ability to quickly discriminate nonorthogonal states and thereby solve unstructured search is a generic feature of nonlinear quantum mechanics.
\end{abstract}

\section{Introduction}
\label{sec:intro}

Linearity is an essential feature of quantum mechanics whose violation can have dramatic operational consequences.  In particular, Abrams and Lloyd \cite{AL98} showed that in a model of nonlinear quantum mechanics due to Weinberg \cite{Wei89}, one can exponentially increase the angle between quantum states, distinguishing states separated by an angle $\epsilon$ in time $O(\log\frac{1}{\epsilon})$.

This ability has strong computational implications, including a fast algorithm for unstructured search (and thus a fast algorithm for any problem in NP).  In the unstructured search problem, one aims to find a marked item using a black box that determines whether any given item (out of $N$ possible items) is marked.  Using the ability to distinguish nonorthogonal states in the Weinberg theory, Abrams and Lloyd gave an algorithm for unstructured search that uses time $O(\log N)$ and only a single query.

More recent work of Meyer and Wong \cite{MW13} considered the computational power of the Gross-Pitaevskii equation, a nonlinear Schr{\"o}dinger equation that provides an effective description of Bose-Einstein condensates.  Specifically, they considered a discrete version of the Gross-Pitaevskii equation,
\begin{align}
  i \frac{\d}{\d{t}} \braket{x}{\psi} = g |{\braket{x}{\psi}}|^2 \braket{x}{\psi}
  \label{eq:gp}
\end{align}
for a unit vector $\ket{\psi} \in \spn\{\ket{x}\}$, where the coefficient $g$ quantifies the strength of the nonlinearity.  They presented an algorithm for unstructured search that simultaneously applies an oracle Hamiltonian, an input-independent driving Hamiltonian, and the Gross-Pitaevskii nonlinearity.  Their algorithm uses time $O(\min\{\sqrt{N/g},\sqrt{N}\})$, so it can (unsurprisingly) solve the search problem arbitrarily fast using an arbitrarily strong nonlinearity.  However, by considering the resources needed to measure the system at a precisely chosen time, they argued that a reasonable implementation would have complexity $O(N^{1/4})$, giving only a modest improvement over the $O(\sqrt{N})$ complexity of Grover's algorithm \cite{Gro97}.

In this paper, we apply the original approach of Abrams and Lloyd to the Gross-Pitaevskii nonlinearity.  In \sec{statedisc}, we find the optimal protocol for distinguishing two states of a qubit by adding a carefully-chosen driving term.  The optimal procedure distinguishes states with overlap $1-\epsilon$ in time $O(\frac{1}{g}\log\frac{1}{\epsilon})$, recovering essentially the same behavior as the Weinberg model.  In \sec{search}, we apply this result to the unstructured search problem.  We present an algorithm with complexity $O(\min\{\frac{1}{g}\log(gN),\sqrt{N}\})$, exponentially improving the upper bound of Meyer and Wong.  Furthermore, we give a lower bound of $\Omega(\min\{\frac{1}{g},\sqrt{N}\})$, showing that our algorithm is optimal up to a logarithmic factor.

It would be interesting to find a variant of quantum theory that allows a polynomial speedup over quantum computation, but no more (see \cite{ABFL14} for some recent progress in this direction).  However, our results suggest that the Gross-Pitaevskii model does not provide such an example.  We also consider a broad class of nonlinearities that generalize the Gross-Pitaevskii model and show that all such models lead to essentially the same behavior.  This supports the view that the ability to exponentially increase the angle between states is a generic feature of nonlinear quantum mechanics, as previously suggested by Abrams and Lloyd \cite{AL98} and Aaronson \cite{Aar05}.

In light of the dramatic consequences for information processing, it seems unlikely that quantum mechanics is fundamentally nonlinear.  However, information-processing protocols using the Gross-Pitaevskii model can be used to place limitations on the validity of the underlying approximation \cite{MW13}.  We discuss this point further in \sec{gpapprox}, where we use our state discrimination protocol to show that for an $\natoms$-atom Bose-Einstein condensate, the Gross-Pitaevskii approximation can only hold up to time $O(\frac{1}{g}\log\natoms)$.

Finally, we conclude with an open problem regarding higher-dimensional nonlinear state discrimination in \sec{conclusion}.

\section{State discrimination}
\label{sec:statedisc}

In nonlinear quantum mechanics, states that are initially nonorthogonal can evolve to become perfectly distinguishable.  In this section, we analyze the optimal protocol for distinguishing two states of a qubit under the Gross-Pitaevskii and related nonlinearities.

Specifically, we consider nonlinear dynamics governed by an equation
\begin{align}
  i \frac{\d}{\d{t}} \ket{\psi} = H(t) \ket{\psi} + K \ket{\psi}
\end{align}
where $H(t)$ is a (time-dependent) Hermitian operator and $K$ is a nonlinearity of the form
\begin{align}
  \bra{x} (K \ket{\psi}) 
  &= \nlin(|{\braket{x}{\psi}}|) \braket{x}{\psi}
\label{eq:nlin}
\end{align} 
where $\nlin\colon[0,1] \to \R$ is a function characterizing the nonlinearity.  For example, the Gross-Pitaevskii nonlinearity corresponds to $\nlin(x)=g x^2$, where the coefficient $g$ quantifies the strength of the nonlinearity.

Suppose we are given one of two possible states of a qubit and our goal is to distinguish them as quickly as possible.  Using the ability to choose the Hermitian driving term $H(t)$, we can orient the states however we like, provided we preserve their inner product.  Thus, to optimally distinguish the given states, we should determine how to orient them on the Bloch sphere so their inner product decreases as quickly as possible.

Consider the pure state with Bloch sphere coordinates $(x,y,z)$, i.e., with density matrix 
\begin{align}
  \rho = \frac{1}{2}\begin{pmatrix} 1+z & x-iy \\ x+iy & 1-z \end{pmatrix}.
\end{align}
Since $|{\braket{0}{\psi}}|^2 = \frac{1+z}{2}$ and $|{\braket{1}{\psi}}|^2 = \frac{1-z}{2}$, the nonlinear term alone is equivalent to the state-dependent Hamiltonian
\begin{align}
  \begin{pmatrix}
    \nlin\bigl((\frac{1+z}{2})^{1/2}\bigr) & 0 \\
    0 & \nlin\bigl((\frac{1-z}{2})^{1/2}\bigr)
  \end{pmatrix}.
\end{align}
In other words, the state evolves according to the equation
\begin{align}
  \frac{\d{\rho}}{\d{t}}
  &= \frac{i}{2} \left[ \begin{pmatrix} 1+z & x-iy \\ x+iy & 1-z \end{pmatrix},
  \begin{pmatrix}
    \nlin\bigl((\frac{1+z}{2})^{1/2}\bigr) & 0 \\
    0 & \nlin\bigl((\frac{1-z}{2})^{1/2}\bigr)
  \end{pmatrix}
  \right] \\
  &= \frac{\nlinred(z)}{2} \begin{pmatrix}0 & -ix-y \\ ix-y & 0\end{pmatrix}
\end{align}
where the odd function $\nlinred(z)\colon [-1,1] \to \R$ is defined by
\begin{align}
  \nlinred(z) := \nlin\bigl((\tfrac{1+z}{2})^{1/2}\bigr) - \nlin\bigl((\tfrac{1-z}{2})^{1/2}\bigr).
  \label{eq:nlinred}
\end{align}
Thus we find
\begin{align}
  \frac{\d}{\d{t}}(x,y,z) &= \nlinred(z) \, (-y, x, 0).
\end{align}
Under these dynamics, states rotate around lines of latitude on the Bloch sphere at a rate depending on their latitude.

\begin{figure}
  \begin{subfigure}{.5\linewidth}
    \centering
  \begin{overpic}[width=2.5in]{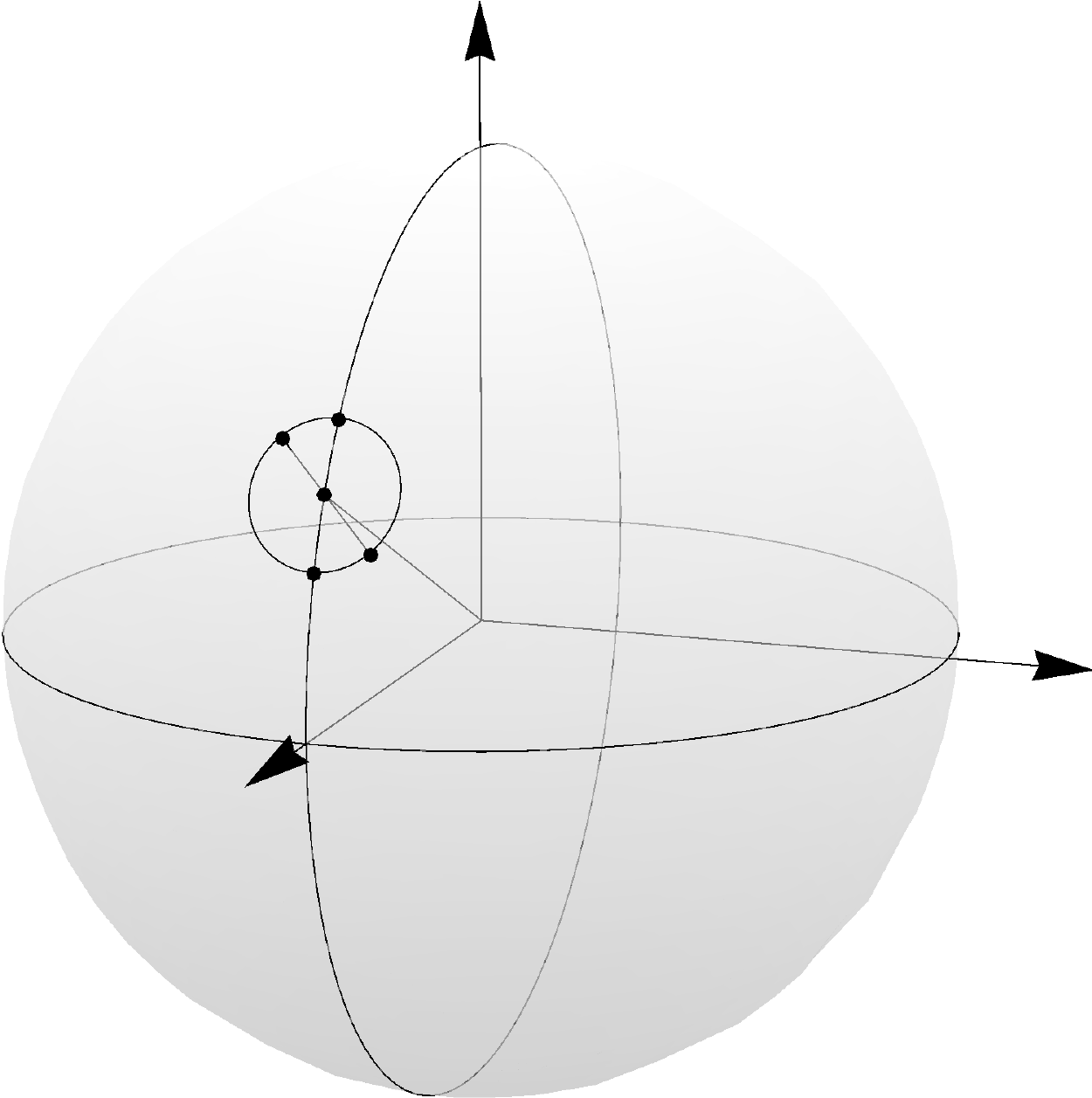}
    \put(18,24){$x$}
    \put(101.5,38){$y$}
    \put(42,102.5){$z$}
    \put(40.2,47.2){\small$\phi$}
    \put(26.7,62.2){\small$\theta$}
    \put(22.25,51.25){\scalebox{.9}{\rotatebox{37}{$\left\{\rule{0pt}{.481cm}\right.$}}}
    \put(22.5,50){\small$\alpha$}
  \end{overpic}
  \caption{\label{fig:states}}
  \end{subfigure}
  \begin{subfigure}{.5\linewidth}
    \centering
  \begin{overpic}[width=2.5in]{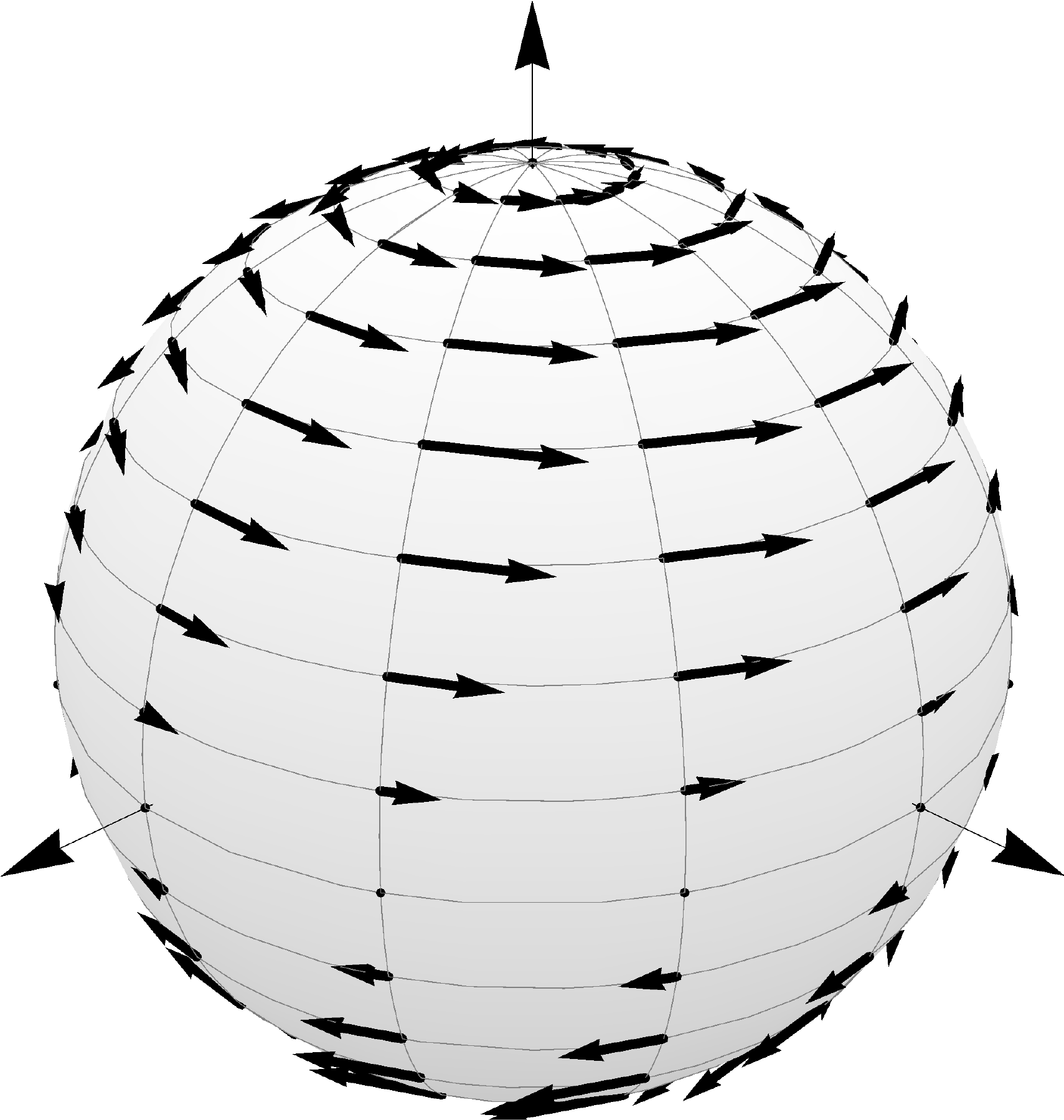}
    \put(-4.5,20){$x$}
    \put(96.5,20){$y$}
    \put(46,102){$z$}
  \end{overpic}
  \caption{\label{fig:gpbloch}}
  \end{subfigure}
\caption{\label{fig:bloch}
\subfig{states} Orientation of states on the Bloch sphere with overlap $\cos\frac{\alpha}{2}$.
\subfig{gpbloch} The flow on the Bloch sphere induced by the Gross-Pitaevskii nonlinearity.}
\end{figure}
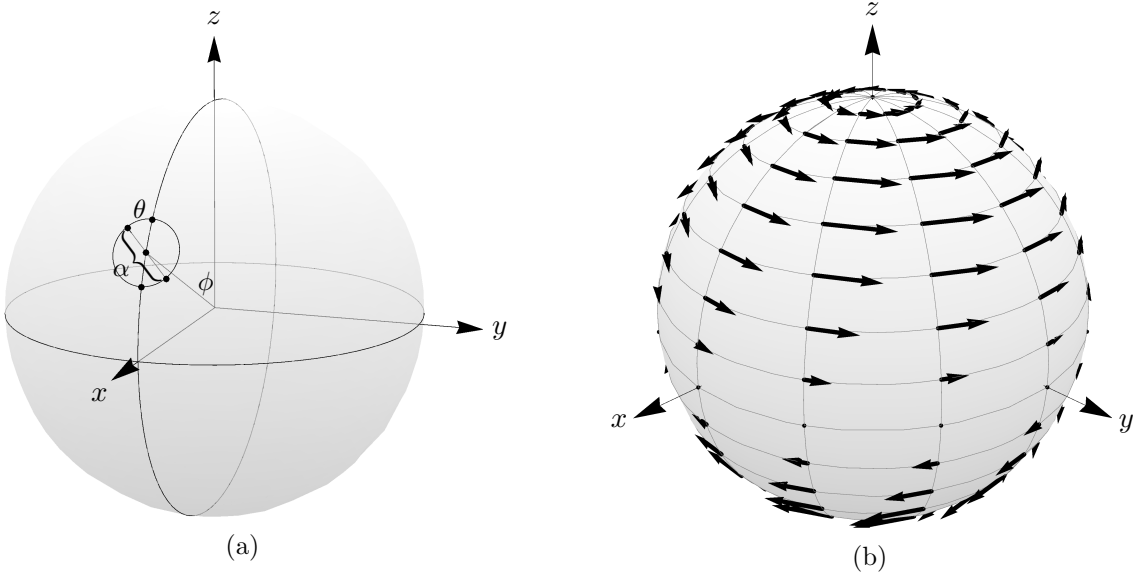

Now consider how to optimally orient two states on the Bloch sphere.  The rate of change of the inner product of Bloch vectors $(x_+,y_+,z_+)$ and $(x_-,y_-,z_-)$ is
\begin{align}
  \frac{\d}{\d{t}}(x_+ x_- + y_+ y_- + z_+ z_-)
  &= (x_+ y_- - y_+ x_-)\bigl(\nlinred(z_+) - \nlinred(z_-)\bigr).
\end{align}
Suppose the states are separated by a fixed angle $\alpha$ on the Bloch sphere (i.e., angle $\alpha/2$ in Hilbert space) and we aim to rotate them to maximize the rate of decrease of their inner product.  Rotations about the $z$ axis do not affect this rate.  Thus, without loss of generality, we can choose the midpoint between the two states to lie in the $xz$ plane.  We orient the states as shown in \fig{states}, where $\phi$ is the polar angle from the positive $z$ axis to the midpoint and $\theta$ is the angle of rotation about the midpoint, with $\theta=0$ corresponding to the states lying along the line of longitude that passes through the $x$ axis.  In terms of these parameters, the states have the form
\begin{align}
  x_\pm &= \cos\tfrac{\alpha}{2}\sin\phi 
           \pm \sin\tfrac{\alpha}{2}\cos\phi\cos\theta \\
  y_\pm &= \pm \sin\tfrac{\alpha}{2}\sin\theta \\
  z_\pm &= \cos\tfrac{\alpha}{2}\cos\phi 
           \mp \sin\tfrac{\alpha}{2}\sin\phi\cos\theta.
\end{align}
Thus the rate of change of the inner product on the Bloch sphere is
\begin{align}
  \frac{\d}{\d{t}}\cos\alpha
  &= (x_+ y_- - y_+ x_-)\bigl(\nlinred(z_+) - \nlinred(z_-)\bigr) \\
  &= \sin\alpha \sin\phi \sin\theta \bigl(\nlinred(z_-) - \nlinred(z_+)\bigr).
\label{eq:iprate}
\end{align}

Given a specific nonlinearity $\nlin$, our goal is to choose $\phi$ and $\theta$ to minimize \eq{iprate}.  Next we perform this calculation for several examples.

\subsection{Gross-Pitaevskii nonlinearity}

For the Gross-Pitaevskii nonlinearity, we have $\nlin(x)=gx^2$, so $\nlinred(z) = gz$ and the states evolve as
\begin{align}
	\frac{\d}{\d{t}}(x,y,z) = gz(-y,x,0).
\label{eq:gpbloch}
\end{align}
As depicted in \fig{gpbloch}, we can view this as a flow on the Bloch sphere that pushes states along lines of latitude.  The rate of rotation varies as a function of latitude, with no rotation at the equator or poles and opposite directions of rotation in the northern and southern hemispheres.

Given a fixed angle between states, we aim to place those states somewhere on the Bloch sphere to maximize the rate at which they separate.  Equation \eq{iprate} shows that the rate of change of the inner product on the Bloch sphere is
\begin{align}
  \frac{\d}{\d{t}}\cos\alpha
  &= g \sin\alpha \sin\tfrac{\alpha}{2} \sin^2\phi \sin2\theta.
\end{align}
Clearly the rate of decrease of $\cos\alpha$ is maximized by choosing $\phi=\pi/2$ and $\theta=3\pi/4$, giving
\begin{align}
  \frac{\d}{\d{t}}\cos\alpha
  &= -g \sin\alpha \sin\tfrac{\alpha}{2}.
\end{align}
Equivalently, the rate of change of the overlap $\cos\frac{\alpha}{2}$ (i.e., the magnitude of the inner product in Hilbert space) is
\begin{align}
  \frac{\d}{\d{t}}\cos\tfrac{\alpha}{2}
  &= \frac{1}{4 \cos\tfrac{\alpha}{2}} \frac{\d}{\d{t}}\cos\alpha \\
  &= -\frac{g}{2}\sin^2\tfrac{\alpha}{2} \\
  &= -\frac{g}{2}(1-\cos^2\tfrac{\alpha}{2}), \label{eq:gpip}
\end{align}
which has the solution
\begin{align}
  \cos\tfrac{\alpha}{2} = \frac{\cos\tfrac{\alpha_0}{2} \cosh \tfrac{gt}{2} - \sinh \tfrac{gt}{2}}{\cosh \tfrac{gt}{2} - \cos\tfrac{\alpha_0}{2} \sinh \tfrac{gt}{2}}
\label{eq:ip}
\end{align}
where $\alpha_0$ is the value of $\alpha$ at $t=0$.  The states become orthogonal in a time $t_\perp$ such that $\tanh \tfrac{gt_\perp}{2} = \cos\tfrac{\alpha_0}{2}$, i.e.,
\begin{align}
  t_\perp = \frac{2}{g} \tanh^{-1}(\cos \tfrac{\alpha_0}{2}) 
          = \frac{2}{g} \ln(\cot\tfrac{\alpha_0}{4}).
\label{eq:disttime}
\end{align}
In particular, if the initial overlap is $\cos\frac{\alpha_0}{2} = 1-\epsilon$, the time to distinguish the states is $\Theta(\log\frac{1}{\epsilon})$.

In their optimal orientation, the states have Bloch vectors
\begin{align}
	x_\pm &= \cos\tfrac{\alpha}{2} \\
	y_\pm &= \pm \tfrac{1}{\sqrt2} \sin\tfrac{\alpha}{2} \\
	z_\pm &= \pm \tfrac{1}{\sqrt2} \sin\tfrac{\alpha}{2}.
\end{align}
To keep the states in this orientation, we must apply a rotation about the $x$ axis that keeps the $y$ and $z$ components equal.  An $x$ rotation (generated by the Hamiltonian $\frac{\omega}{2}\sigma_x$) gives $\frac{\d}{\d{t}}(x,y,z) = \omega(0,-z,y)$.  Combining this with the effect of the nonlinearity in \eq{gpbloch}, we have 
\begin{align}
  \frac{\d}{\d{t}}(x,y,z) = (-gyz,gxz-\omega z,\omega y).
\label{eq:blochdynamics}
\end{align}
The states remain optimally oriented when
\begin{align}
  0
  &= \frac{\d}{\d{t}}(y-z) \\
  &= gxz - \omega(y+z),
\end{align}
so we must choose
\begin{align}
	\frac{\omega}{g}
	&= \frac{xz}{y+z} \\
	&= \tfrac{1}{2} \cos\tfrac{\alpha}{2}
\label{eq:omega}
\end{align}
(which is given as an explicit function of $t$ by \eq{ip}).
One can easily verify that, with this choice of $\omega$, equation \eq{blochdynamics} is satisfied.

\begin{figure}
\begin{subfigure}{.5\linewidth}
  \centering
  \begin{overpic}[height=1.85in]{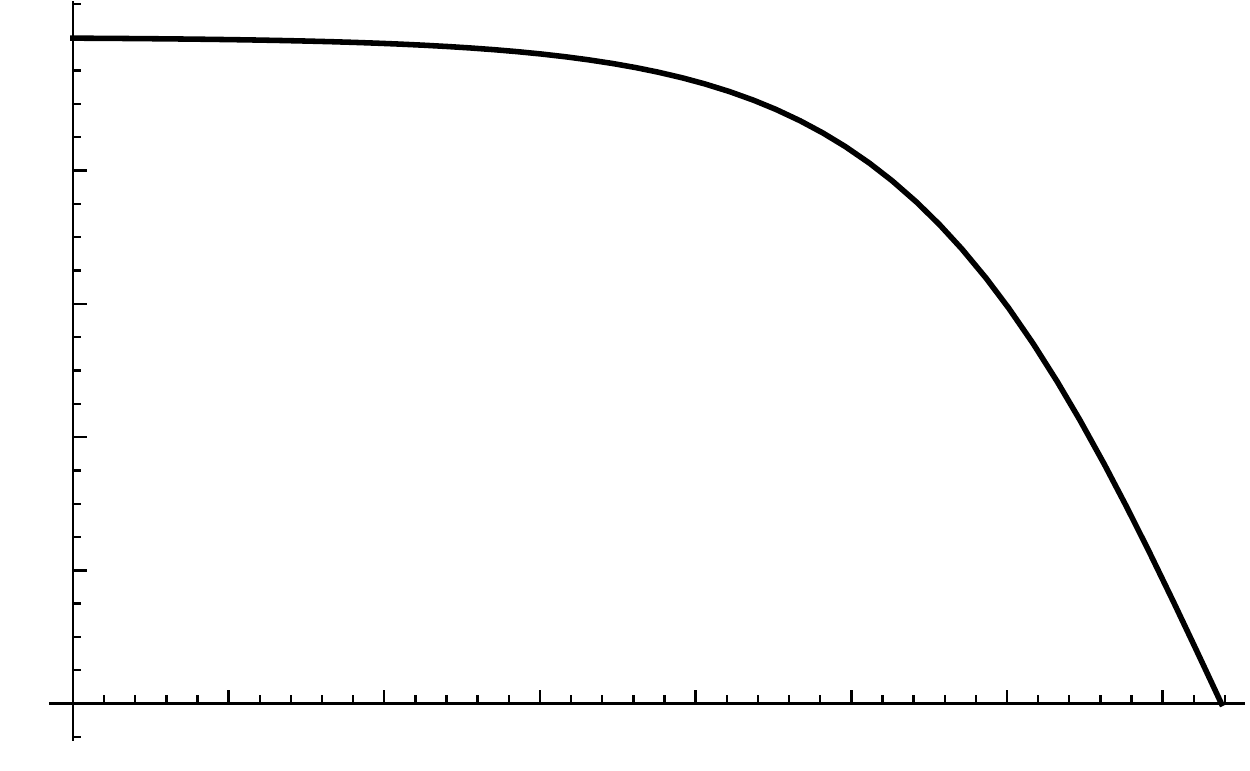}
    \put(101,4){$gt$}
    \put(.5,64){$\cos\tfrac{\alpha}{2}$}
    \put(17.2,1){\footnotesize$1$}
    \put(29.5,1){\footnotesize$2$}
    \put(42,1){\footnotesize$3$}
    \put(54.4,1){\footnotesize$4$}
    \put(66.9,1){\footnotesize$5$}
    \put(79.3,1){\footnotesize$6$}
    \put(91.7,1){\footnotesize$7$}
    \put(-1,15){\footnotesize$0.2$}
    \put(-1,25){\footnotesize$0.4$}
    \put(-1,36){\footnotesize$0.6$}
    \put(-1,47){\footnotesize$0.8$}
    \put(-1,57){\footnotesize$1.0$}
  \end{overpic}
\caption{\label{fig:ip}}
\end{subfigure}
\begin{subfigure}{.5\linewidth}
  \centering
  \begin{overpic}[height=1.85in]{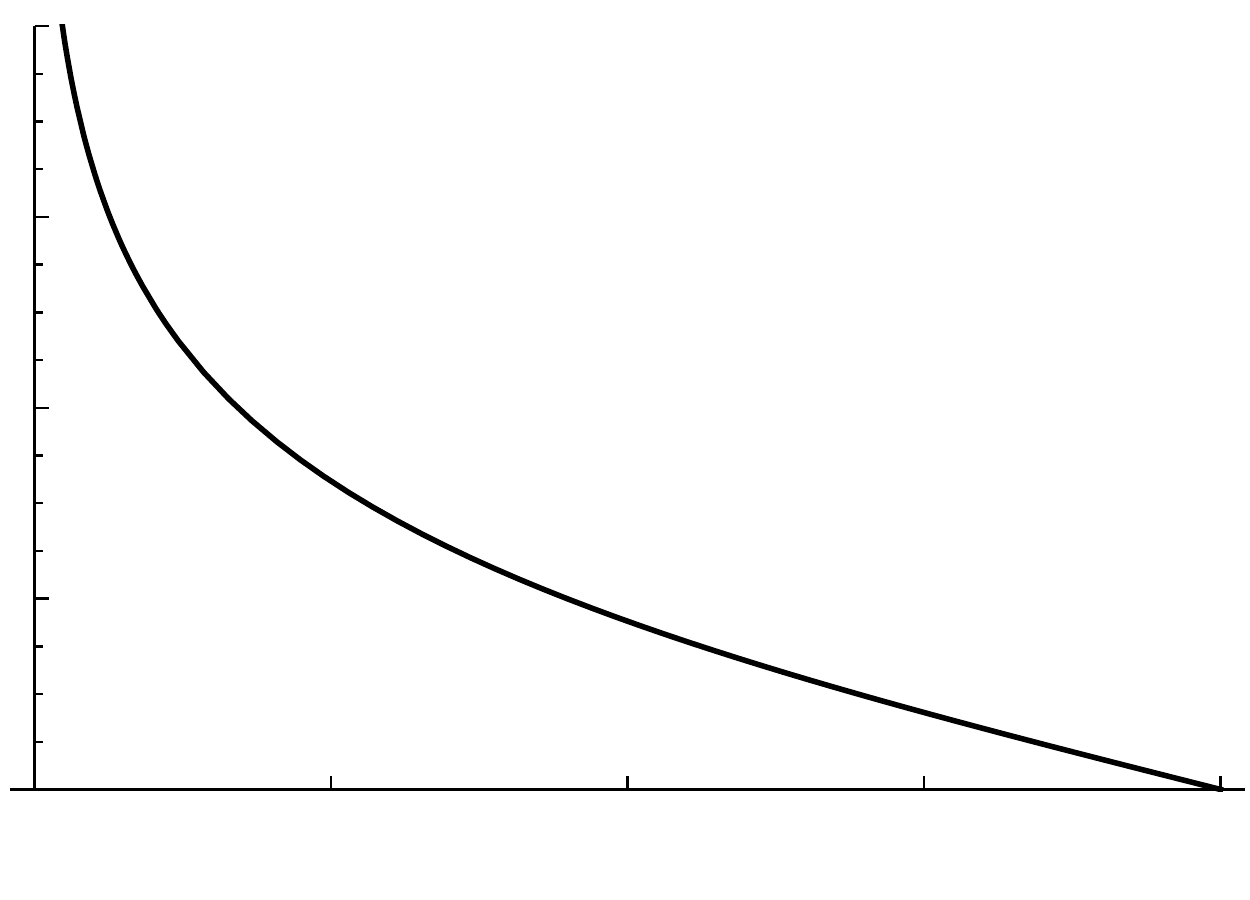}
    \put(101,8){$\alpha_0$}
    \put(0,74){$gt_\perp$}
    \put(1.6,4){\footnotesize$0$}
    \put(24.8,4){\footnotesize$\tfrac{\pi}{4}$}
    \put(48.5,4){\footnotesize$\tfrac{\pi}{2}$}
    \put(71.5,4){\footnotesize$\tfrac{3\pi}{4}$}
    \put(96.2,4){\footnotesize$\pi$}
    \put(-1,22.75){\footnotesize$2$}
    \put(-1,38){\footnotesize$4$}
    \put(-1,53.25){\footnotesize$6$}
    \put(-1,68.5){\footnotesize$8$}
  \end{overpic}
\caption{\label{fig:t}}
\end{subfigure}
\caption{\label{fig:gp}Optimally distinguishing two states of a qubit using the Gross-Pitaevskii nonlinearity $\nlin(x)=gx^2$.
\subfig{ip} Inner product between states as a function of time, as given by equation \eq{ip}, for initial states with $\alpha_0=0.1$.
\subfig{t} Time to distinguish states separated by an angle $\alpha_0$ on the Bloch sphere, as given by equation \eq{disttime}.}
\end{figure}
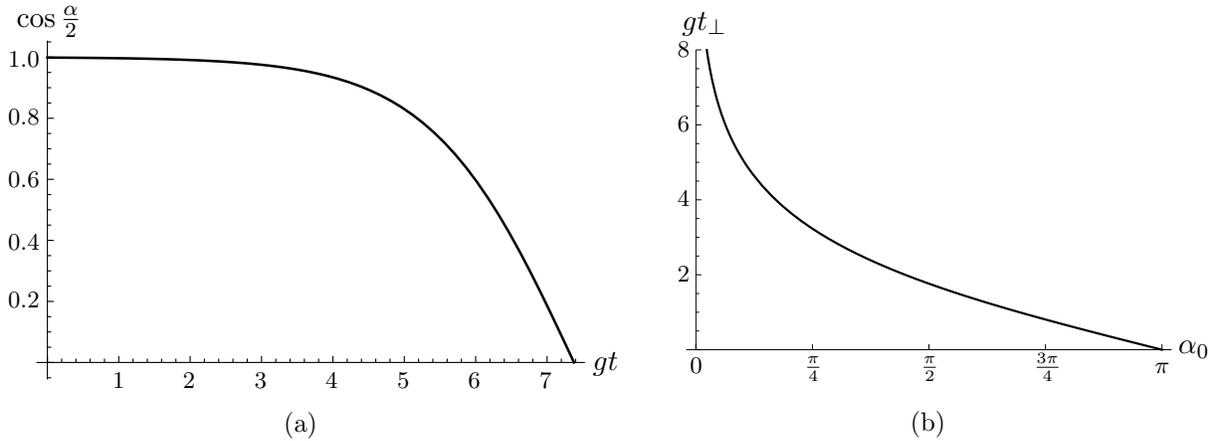

The performance of the optimal procedure using the Gross-Pitaevskii nonlinearity is illustrated in \fig{gp}.  As shown in \fig{ip}, if the states are initially close, they separate gradually at first, with an accelerating rate of separation, until they become orthogonal.  The time for the states to become orthogonal (as a function of their initial separation) is plotted in \fig{t}.

\subsection{Logarithmic nonlinearity}

Motivated by a connection to Bose liquids, reference \cite{MW14} considers the nonlinearity $\nlin(x) = g\ln(x^2)$.  With this nonlinearity, we have $\nlinred(z) = g\ln\frac{1+z}{1-z}$, so by \eq{iprate}, the inner product on the Bloch sphere evolves as
\begin{align}
	\frac{\d}{\d{t}} \cos\alpha 
	&= g \sin\alpha \sin\phi \sin\theta \ln\frac{(1+z_-)(1-z_+)}{(1-z_-)(1+z_+)}.
\end{align}

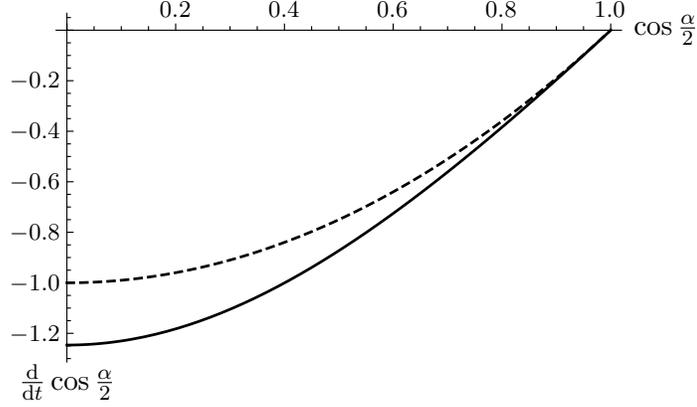
\begin{figure}
\begin{center}
	\raisebox{2ex}{
  \begin{overpic}[height=1.85in]{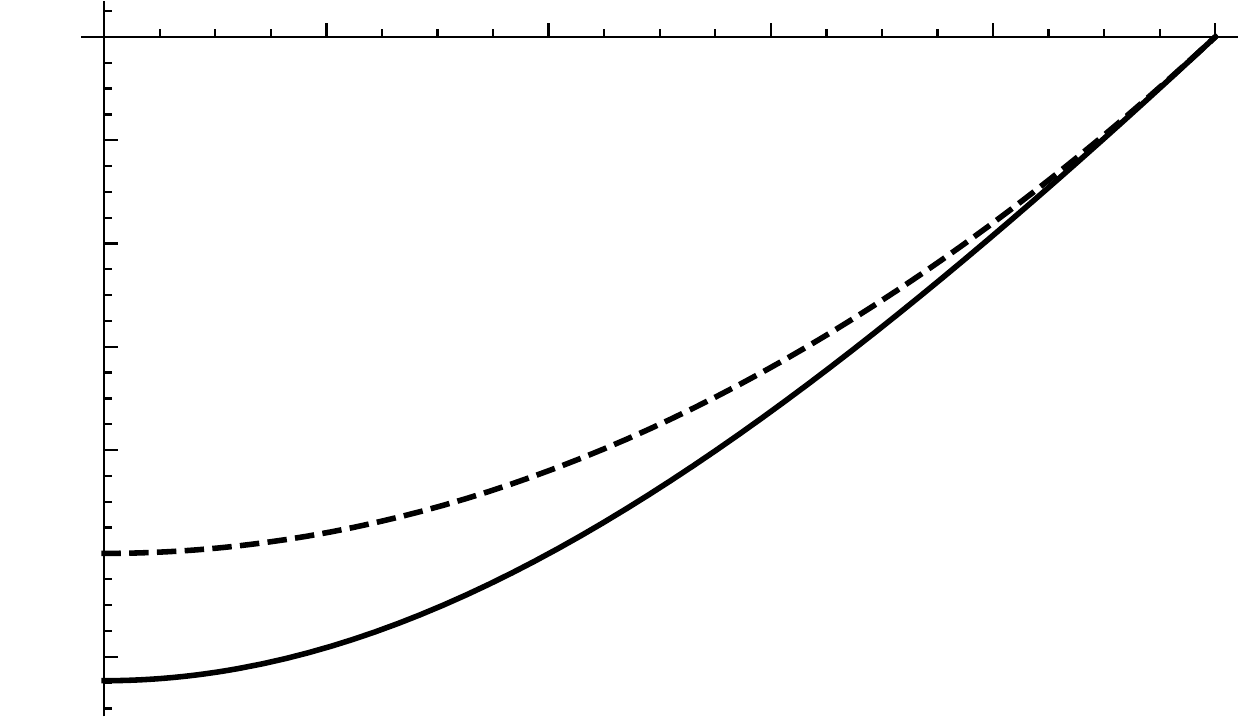}
    \put(101,54){$\cos\tfrac{\alpha}{2}$}
    \put(.5,-4){$\tfrac{\d}{\d{t}}\cos\tfrac{\alpha}{2}$}
    \put(23.5,57){\footnotesize$0.2$}
    \put(41.25,57){\footnotesize$0.4$}
    \put(59,57){\footnotesize$0.6$}
    \put(76.75,57){\footnotesize$0.8$}
    \put(94.5,57){\footnotesize$1.0$}
    \put(-1,45.5){\footnotesize$-0.2$}
    \put(-1,37.2){\footnotesize$-0.4$}
    \put(-1,28.9){\footnotesize$-0.6$}
    \put(-1,20.6){\footnotesize$-0.8$}
    \put(-1,12.3){\footnotesize$-1.0$}
    \put(-1,4){\footnotesize$-1.2$}
  \end{overpic}}
\end{center}
\caption{\label{fig:log}Rate of change of the overlap for the Gross-Pitaevskii and logarithmic nonlinearities.  The solid curve is \eq{logip} with $g=1$; the dashed curve is \eq{gpip} with $g=2$.}
\end{figure}

Consider the initial states that are optimal for the Gross-Pitaevskii nonlinearity, with $\phi=\pi/2$ and $\theta=3\pi/4$.  While this choice is suboptimal for the logarithmic nonlinearity, it serves to place an upper bound on the time required to distinguish the states.  (In fact, numerical calculation shows that the optimal states have $\phi=\pi/2$ and $\theta \approx 3\pi/4$.)  Using these states, we have
\begin{align}
  \frac{\d}{\d{t}} \cos\tfrac{\alpha}{2}
  &= g \frac{1}{\sqrt{2}} \ln\biggl(\frac{\sqrt{2}-\sin\frac{\alpha}{2}}{\sqrt{2}+\sin\frac{\alpha}{2}}\biggr) \sin\tfrac{\alpha}{2}.
  \label{eq:logip}
\end{align}
As shown in \fig{log}, this is at most a constant times the corresponding quantity under the Gross-Pitaevskii nonlinearity, so the performance of the logarithmic nonlinearity is qualitatively similar.  In particular, for $\cos\frac{\alpha_0}{2}=1-\epsilon$ we again find that the time to distinguish the states is $\Theta(\log\frac{1}{\epsilon})$.

\subsection{General nonlinearities}
\label{sec:general}

So far, we have considered the performance of two specific nonlinearities.  However, similar considerations apply for a wide class of nonlinearities of the form \eq{nlin}.

Since the evolution of states under a nonlinearity $\nlin$ depends only on the function $\nlinred$ defined in \eq{nlinred}, let us evaluate state discrimination in terms of the latter function.  Note that we can achieve any odd function $\nlinred$ for some $\nlin$.  In particular, if we let
\begin{align}
  \nlin(x) = \begin{cases}
    \mu(x) & x \in [0,\tfrac{1}{\sqrt{2}}] \\
    \nu(\sqrt{1-x^2}) & x \in (\tfrac{1}{\sqrt{2}},1]
  \end{cases}
\end{align}
for some $\mu,\nu\colon [0,\frac{1}{\sqrt{2}}] \to \R$, then we find
\begin{align}
  \nlinred(z) = \begin{cases}
    \nu\bigl(\sqrt{\tfrac{1-z}{2}}\bigr) - \mu\bigl(\sqrt{\tfrac{1-z}{2}}\bigr)
    & z \in (0,1] \\
    0 & z=0 \\
    -\nlinred(-z) & z \in [-1,0).
  \end{cases}
\end{align}

Now suppose $\nlinred(z)$ is approximately linear for small $z$.  Specifically, suppose there are constants $g,\delta>0$ such that $\nlinred(z) \ge gz$ for all $z \in [0,\delta]$.  For such a nonlinearity, the complexity of distinguishing states with overlap $1-\epsilon$ is $O(\frac{1}{g}\log\frac{1}{\epsilon})$, just as with the Gross-Pitaevskii and logarithmic nonlinearities.

To see this, a straightforward calculation shows that the states separated by an angle $\alpha$ with $\phi=\pi/2$ and $\theta = 3\pi/4$ have
\begin{align}
  \frac{\d}{\d{t}}\cos\tfrac{\alpha}{2} 
  = -\frac{1}{\sqrt2} \nlinred\bigl(\tfrac{1}{\sqrt2}\sin\tfrac{\alpha}{2}\bigr) \sin\tfrac{\alpha}{2}.
  \label{eq:generalip}
\end{align}
Under the given conditions on $\nlinred(z)$, we find
\begin{align}
  \frac{\d}{\d{t}}\cos\tfrac{\alpha}{2} 
  \le -\frac{g}{2} \sin^2\tfrac{\alpha}{2}
\end{align}
provided $\sin\tfrac{\alpha}{2} \le \sqrt{2}\delta$.  Comparing with \eq{gpip}, we see that the states separate at least as quickly as with the Gross-Pitaevskii nonlinearity provided the overlap is at least $\sqrt{1-2\delta^2}$.  Similarly to \eq{disttime}, a straightforward calculation shows that the time to achieve overlap $\cos\frac{\alpha}{2} \ge \sqrt{1-2\delta^2}$ under the Gross-Pitaevskii nonlinearity is at most
\begin{align}
  \frac{2}{g}\ln\biggl(\frac{\cot\tfrac{\alpha_0}{4}}{\cot\tfrac{\alpha}{4}}\biggr),
\end{align}
so the states become distinguishable with constant advantage in time $O(\frac{1}{g}\log\frac{1}{\epsilon})$.

In fact, it is not essential for $\nlinred(z)$ to be linear about $z=0$; similar performance can be achieved provided the function grows linearly about any fixed $z_0 \in [0,1)$ (which holds for any function that is differentiable and non-constant over some interval).  Specifically, suppose there are constants $g,\Delta>0$ such that $|\nlinred(z_0) - \nlinred(z_0+\delta)| \ge g|\delta|$ for all $|\delta|<\Delta$.  Then, taking $\cos\phi=z_0$ and either $\theta=3\pi/4$ or $\theta=\pi/4$ (making the choice so that $\nlinred(z_+) > \nlinred(z_-)$), we find
\begin{align}
  |z_+-z_-|
  &= \sqrt{2(1-z_0^2)}\sin\tfrac{\alpha}{2} \\
  &\ge \sqrt{\frac{1-z_0^2}{2}} \, \alpha.
\end{align}
Thus by \eq{iprate}, we have
\begin{align}
  \frac{\d}{\d{t}}\cos\alpha
  \le -g \sqrt{\frac{1-z_0^2}{2}} \, \alpha \sin\alpha
\end{align}
provided $\alpha$ is at most some constant (depending on $z_0$ and $\Delta$).  Integrating this inequality, we have $\alpha(t) \ge e^{ct} \alpha_0$ where $c := g \sqrt{(1-z_0^2)/2}$.  Since the angle between the states increases exponentially until it reaches at least some constant, states with overlap $1-\epsilon$ become distinguishable with constant advantage in time $O(\frac{1}{g} \log\frac{1}{\epsilon})$.

Conversely, provided only that $\nlinred(z)$ is Lipschitz continuous, this behavior is asymptotically optimal.  To see this, suppose that $\nlinred(z)$ has Lipschitz constant $g$, i.e., $|\nlinred(z)-\nlinred(z+\delta)| \le g\delta$.  Since
\begin{align}
  |z_+ - z_-|^2
  &\le \|(x_+,y_+,z_+) - (x_-,y_-,x_-) \|^2 \\
  &= 2(1-\cos\alpha) \\
  &\le \alpha^2,
\end{align}
we have $|\nlinred(z_-) - \nlinred(z_+)| \le 2 g \alpha$ in \eq{iprate}.  Therefore
\begin{align}
  \frac{\d}{\d{t}}\cos\alpha
  \ge -2g \alpha \sin\alpha.
\end{align}
Integrating this inequality gives $\alpha(t) \le e^{2g t} \alpha_0$.  Since the angle increases at most exponentially with $gt$, we require time $\Omega(\frac{1}{g}\log\frac{1}{\epsilon})$ to distinguish states with overlap $1-\epsilon$.

Note that it is possible to violate this lower bound if the Lipschitz condition does not hold.  For example, suppose $\nlinred(z) = \sgn(z) \, \sqrt{|z|}$ (which is achieved, for example, with $\nlin(x)=0$ for $x \in [0,{1}/{\sqrt2}]$ and $\nlin(x)=\sqrt{2x^2-1}$ for $x \in ({1}/{\sqrt2},1]$).  Then, taking states with $\phi=\pi/2$ and $\theta=3\pi/4$, equation \eq{generalip} gives
\begin{align}
  \frac{\d}{\d{t}}\cos\tfrac{\alpha}{2}
  = - \biggl(\frac{1}{\sqrt2} \sin\tfrac{\alpha}{2}\biggr)^{3/2}.
\end{align}
While this differential equation does not have a simple closed-form solution, we have the bound
\begin{align}
  \frac{\d}{\d{t}}\cos\tfrac{\alpha}{2}
  \le - r \sqrt{\alpha} \sin\tfrac{\alpha}{2}
\end{align}
for all $\alpha \in [0,\pi]$, where $r := (2^{3/4}\sqrt\pi)^{-1}$.  Integrating this inequality, we find $\alpha(t) \ge (\sqrt{\alpha_0}+rt)^2$.  Therefore any two distinct states become distinguishable in constant time, independent of how close they are initially.

\section{Unstructured search}
\label{sec:search}

We now turn our attention to algorithms for unstructured search in nonlinear quantum mechanics.  In the unstructured search problem, our goal is to search the set $[N] := \{1,2,\ldots,N\}$ for a member of the marked set $M \subseteq [N]$, given the ability to determine whether a given element is marked.  Equivalently, we can consider the problem of deciding whether there is no marked item (i.e., $M$ is empty) or a unique marked item ($M = \{m\}$ for some unknown $m \in [N]$).  Using standard techniques, an algorithm for this decision problem can be used to find a marked item with only logarithmic overhead.

In the conventional quantum query model, access to the input is provided by a black box that determines whether a given element of $[N]$ is marked (and that can be queried on a superposition of elements).  Here we work in a continuous-time model \cite{FG98} where access to the input is provided by an oracle Hamiltonian.  This Hamiltonian is $\ket{m}\bra{m}$ if $M=\{m\}$ and is zero if there is no marked item.  We consider algorithms that use such an oracle Hamiltonian together with an arbitrary $M$-independent driving Hamiltonian $H(t)$ and a given nonlinearity $K$.  The dynamics of such an algorithm are governed by the equation
\begin{align}
  i \frac{\d}{\d{t}} \ket{\psi} = \bigl(\ket{m}\bra{m} + H(t)\bigr) \ket{\psi} + K \ket{\psi}
\label{eq:algham-marked}
\end{align}
if $M=\{m\}$, or
\begin{align}
  i \frac{\d}{\d{t}} \ket{\psi} = H(t) \ket{\psi} + K \ket{\psi}
\label{eq:algham-nomarked}
\end{align}
if there is no marked item.

We quantify the complexity of such an algorithm by the time required to find the solution.  As in the standard continuous-time query model, we place no constraints on the norm of $H(t)$.  However, we consider a fixed-strength nonlinearity $K$, since an arbitrarily strong nonlinearity could solve the search problem arbitrarily fast.  Thus our complexities will be expressed as a function of both $N$, the number of items, and $g$, a parameter characterizing the strength of the nonlinearity.  In contrast, in the absence of a nonlinearity, the search problem requires $\Omega(\sqrt N)$ queries even when $H(t)$ can be arbitrarily large \cite{FG98}.

\subsection{Algorithm}
\label{sec:alg}

\begin{figure}
  \[
  \Qcircuit @C=1em @R=.7em @!R {
  \lstick{\ket{0}} & \gate{H} & \ctrl{1} & \gate{H} & \qw \\
  \lstick{\ket{s}} & \qw & \gate{U} & \qw & \qw
  }\]
  \caption{\label{fig:hadamard}The Hadamard test.  The first register stores a qubit.  The Hadamard gate $H$ acts as $H\ket{0} = \frac{1}{\sqrt2}(\ket{0}+\ket{1})$ and $H\ket{1} = \frac{1}{\sqrt2}(\ket{0}-\ket{1})$).}
\end{figure}
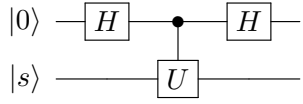

To give an algorithm for unstructured search, we reduce the problem of deciding whether there are zero or one marked items to the task of discriminating two possible states of a qubit.  We do this using the Hadamard test shown in \fig{hadamard}.  (This approach is similar to that of Abrams and Lloyd \cite{AL98}, although we work with the continuous-time query model so we can consider algorithms that query the oracle for only a short time.)  A straightforward calculation shows that the output of the Hadamard test circuit is
\begin{align}
  \frac{1}{2}[ \ket{0} (\ket{s} + U\ket{s}) + \ket{1} (\ket{s} - U \ket{s})].
\end{align}
To produce a single-qubit state, we postselect the second register on the state $\ket{s}$.  This postselection succeeds with probability
\begin{align}
  \frac{1}{4}\bigl[|{1 + \bra{s}U\ket{s}}|^2 + |{1 - \bra{s}U\ket{s}}|^2\bigr]
  &= \frac{1+|{\bra{s}U\ket{s}}|^2}{2}
  \label{eq:psprob}
\end{align}
and results in a state
\begin{align}
  \frac{(1 + \bra{s}U\ket{s}) \ket{0} + (1 - \bra{s}U\ket{s}) \ket{1}}
       {\sqrt{2(1+|{\bra{s}U\ket{s}}|^2)}}.
  \label{eq:psstate}
\end{align}

We apply this procedure to the evolution under the oracle Hamiltonian for time $t_1$, namely $U=e^{-i t_1 \ket{s}\bra{s}}$ if element $m$ is marked, or $U=I$ if no item is marked.  We choose the uniform superposition $\ket{s} := \frac{1}{\sqrt{N}}\sum_{x \in [N]}\ket{x}$ as the initial state.

If no item is marked, then clearly $\bra{s}U\ket{s}=1$ for any evolution time $t_1$.  Thus the postselection on $\ket{s}$ always succeeds, and the postselected state of the first qubit is $\ket{0}$.

On the other hand, if some item $m$ is marked, then we have
\begin{align}
  U\ket{s}
  &= \frac{1}{\sqrt N} (e^{-i t_1} \ket{m} + \sum_{x \ne m} \ket{x}).
\end{align}
Therefore $\bra{s}U\ket{s} = 1 - \frac{1}{N} (1-e^{-it_1})$, so the success probability \eq{psprob} is
\begin{align}
  \frac{1+|{\bra{s}U\ket{s}}|^2}{2}
  = 1 - \frac{N-1}{N^2}(1-\cos t_1)
  = 1 - O(t_1^2/N)
\end{align}
(i.e., for $t_1 \ll \sqrt{N}$, the postselection almost always succeeds).
The overlap of the postselected state \eq{psstate} with $\ket{0}$ is
\begin{align}
  \frac{|{1 + \bra{s}U\ket{s}}|}{\sqrt{2(1+|{\bra{s}U\ket{s}}|^2)}}
  &= \frac{|2N-1+e^{-it_1}|}{2\sqrt{N^2 - (N-1)(1-\cos t_1)}} \\
  &= 1 - \frac{t_1^2}{8N^2} + O(t_1^4/N^2).
\label{eq:psip}
\end{align}

It remains to distinguish the two possible states.  Using the protocol described in \sec{statedisc}, states with overlap $1-\epsilon$ can be distinguished in time $t_2 = O(\tfrac{1}{g} \log\tfrac{1}{\epsilon})$, where $g$ is the strength of the nonlinearity.  (We saw in \sec{general} that such a bound holds not only for the Gross-Pitaevskii and logarithmic nonlinearities, but for any nonlinearity of the form \eq{nlin} where $\nlinred$ changes at least linearly over some constant-size interval.)  From \eq{psip}, we have $\epsilon = \Theta(t_1^2/N^2)$.  Thus we find an algorithm that solves the search problem in total time
\begin{align}
  t_1 + t_2 = O(t_1 + \tfrac{1}{g} \log\tfrac{N}{t_1}).
\end{align}
Taking $t_1 = \Theta(\tfrac{1}{g} \log(Ng))$, the total time is $O(\tfrac{1}{g} \log(Ng))$.

Of course, if $g$ is very small (in particular, if $g \ll \frac{\log N}{\sqrt{N}}$) then it may be preferable to eschew the nonlinearity and instead use Grover's algorithm alone.  Taking that possibility into account, we find an algorithm with complexity
\begin{align}
  O(\min\{\tfrac{1}{g} \log(gN),\sqrt{N}\})
  = O\Biggl(\frac{\sqrt{N}}{\frac{g\sqrt{N}}{\log(gN)}+1}\Biggr).
\end{align}
This improves the previous upper bound of $O\bigl(\smash{\sqrt{\frac{N}{g+1}}}\bigr) = O(\min\{\sqrt{N/g},\sqrt{N}\})$ \cite{MW13}.  For example, with $g=\Theta(1)$, we improve the complexity from $O(\sqrt{N})$ to $O(\log N)$.

\subsection{Lower bound}
\label{sec:lb}

We now show that the algorithm described above is nearly optimal.  We follow the same strategy as in the lower bound for the linear case \cite{FG98}.

Let $\ket{\psi}$ be the state of the algorithm when there is no marked item and let $\ket{\psi_m}$ be the state when the marked item is $m$.  Consider how the inner product $\braket{\psi}{\psi_m}$ evolves under the dynamics \eq{algham-marked} and \eq{algham-nomarked} for an arbitrary driving Hamiltonian $H(t)$.  We find
\begin{align}
	\frac{\d}{\d{t}} \braket{\psi}{\psi_m}
	&= -i \braket{\psi}{m}\braket{m}{\psi_m} 
	   +i (K \ket{\psi})^\dag \ket{\psi_m} 
	   -i \bra{\psi} (K \ket{\psi_m}) \\
	&= -i \braket{\psi}{m}\braket{m}{\psi_m} 
	   +i \sum_x \bigl( \nlin(|{\braket{x}{\psi}}|) -\nlin(|{\braket{x}{\psi_m}}|) \bigr)
	    \braket{\psi}{x}\braket{x}{\psi_m}.
\end{align}
(In particular, observe that the driving Hamiltonian $H(t)$ does not appear, just as in the linear case.)  Provided $|\nlin(x)| \le g$, we have
\begin{align}
  \biggl| \sum_x \bigl( \nlin(|{\braket{x}{\psi}}|) -\nlin(|{\braket{x}{\psi_m}}|) \bigr)
  \braket{\psi}{x}\braket{x}{\psi_m} \biggr|
  &\le 2g \sum_x |\braket{\psi}{x}\braket{x}{\psi_m}|
  \le 2g
\end{align}
by the Cauchy-Schwarz inequality, so
\begin{align}
	\Bigl| \frac{\d}{\d{t}} \braket{\psi}{\psi_m} \Bigr|
  &\le |{\braket{m}{\psi}}| + 2g.
\end{align}
Summing over the $N$ possible marked items, we find
\begin{align}
  \frac{\d}{\d{t}} \sum_{m \in [N]} |{\braket{\psi}{\psi_m}}|
  &\le \Biggl(\sum_{m \in [N]} |{\braket{m}{\psi}}|\Biggr) + 2gN \\
  &\le \sqrt{N} + 2gN
\end{align}
where we again used the Cauchy-Schwarz inequality.  Integrating for time $t$ using the initial condition $\sum_{m \in [N]} |{\braket{\psi}{\psi_m}}| = N$, we find that
\begin{align}
  \sum_{m \in [N]} |{\braket{\psi}{\psi_m}}| \ge N - t \sqrt{N}(1+2g\sqrt{N}).
\label{eq:finalip}
\end{align}

For the algorithm to succeed with constant probability, the unmarked state must be distinguishable from each marked state with constant probability, so the final states must satisfy $|{\braket{\psi}{\psi_m}}| \le 1-\delta$ for some constant $\delta > 0$.  Therefore
\begin{align}
	\sum_{m \in [N]} |{\braket{\psi}{\psi_m}}| \le N(1-\delta).
\end{align}
Comparing with \eq{finalip}, we find
\begin{align}
  t \ge \frac{\delta\sqrt{N}}{1+2g\sqrt{N}}.
\end{align}
Thus a bounded-error algorithm must have complexity $\Omega\bigl(\frac{\sqrt{N}}{g\sqrt{N}+1}\bigr) = \Omega(\min\{\frac{1}{g},\sqrt{N}\})$.  This shows that the algorithm described in \sec{alg} is optimal up to a logarithmic factor.

\section{Validity of the Gross-Pitaevskii approximation}
\label{sec:gpapprox}

Since the Gross-Pitaevskii model is an approximate description of a fundamentally quantum system, upper bounds on the search problem in this model can be interpreted as establishing limitations on the effectiveness of the approximation.  Using limits on the parallelizability of unstructured search, Meyer and Wong argued that this perspective gives a lower bound on how large a Bose-Einstein condensate should be for the Gross-Pitaeveskii approximation to hold \cite{MW13}.  They suggested that the $\natoms$-particle Bose-Einstein condensate makes $\natoms$ queries in parallel per unit time, so the bound $\natoms T^2 = \Omega(N)$ for a parallel quantum search algorithm with query depth $T$ \cite{Zal99}, together with their search upper bound $T = O(\sqrt{N/g})$, implies $\natoms = \Omega(g)$.

While the search algorithm presented in \sec{search} has lower complexity, the above argument does not apply to that case.  In our approach, we query the black box for a short time using only linear quantum mechanics and then distinguish the resulting states using the nonlinearity.  Since the nonlinear portion of the computation makes no access to the oracle, the hardness of parallelizing quantum search does not restrict the validity of the Gross-Pitaevskii approximation.

However, we can use state discrimination to obtain a limitation on the time for which the Gross-Pitaevskii approximation remains valid.  In this approximation, $\ket{\psi}$ is not literally the quantum state of the system; rather, it parameterizes the mean-field bosonic state
\begin{align}
  \ket{\MF(\psi)} := \frac{1}{\sqrt{\natoms!}} \biggl( \sum_x \braket{x}{\psi} a_x^\dag \biggr)^\natoms \ket{0}
\end{align}
where $\ket{0}$ is the vacuum state, $\smash{a_x^\dag}$ is a creation operator for mode $x$, and $\natoms$ is the number of particles in the system.
It is straightforward to understand the distinguishability of these mean-field states in terms of their parameterizing vectors: a simple calculation shows that $\braket{\MF(\psi)}{\MF(\phi)} = (\braket{\psi}{\phi})^\natoms$, so the distinguishability of states of the form $\ket{\MF(\psi)}$ is the same as that of $\smash{\ket{\psi}^{\otimes\natoms}}$.
In our protocol, $\ket{\psi}$ is a qubit, so we can represent it using a two-mode condensate and implement the driving term $H(t)$ by driving transitions between these two modes.
By the Helstrom bound \cite{Hel76}, $\natoms = \Theta(1/\epsilon)$ copies are necessary and sufficient to distinguish states with overlap $1-\epsilon$ with constant advantage (in fact, this bound can be achieved even with unentangled measurements \cite{ABBMM05}).  Since the Gross-Pitaevskii nonlinearity of strength $g$ can distinguish states with overlap $1-\epsilon$ in time $t = O(\frac{1}{g}\log\frac{1}{\epsilon})$, we find that the Gross-Pitaevskii approximation can only hold up to time $O(\smash{\frac{1}{g}\log\natoms})$.
For a homogeneous condensate, $g=U\natoms$ for some constant interaction strength $U$, so the approximation only holds up to time $O(\frac{\log\natoms}{\natoms})$.

While it is appealing to obtain such a limitation from information-theoretic considerations, it appears that one can derive a stronger bound by direct analysis of the condensate dynamics \cite{Gor15}.  This analysis suggests that in fact the approximation is only valid for $Ut \ll 1/\natoms$, so that for a homogeneous condensate, it only holds up to time $O(1/\natoms)$.

\section{Conclusion}
\label{sec:conclusion}

We have studied the power of nonlinear quantum dynamics to distinguish nonorthogonal states and to perform unstructured search.  We showed that the Gross-Pitaevskii nonlinearity of strength $g$ can be used to distinguish states with overlap $1-\epsilon$ in time $O(\frac{1}{g}\log\frac{1}{\epsilon})$.  We applied this result to give an algorithm for unstructured search in this model with complexity $O(\min\{\frac{1}{g}\log(gN),\sqrt{N}\})$, an exponential improvement over previous work.  We also showed that many other nonlinearities of a related form have similar behavior.  Finally, we used our state discrimination protocol to argue that for an $\natoms$-atom Bose-Einstein condensate, the Gross-Pitaevskii approximation can only hold up to time $O(\frac{1}{g} \log\natoms)$.

We conclude with an open problem regarding nonlinear state discrimination.  Our state discrimination analysis was restricted to the case of a single qubit.  It is unclear whether states could be discriminated more quickly using higher-dimensional systems.  Numerical evidence suggests that our optimal protocol for distinguishing two states of a qubit with the Gross-Pitaevskii nonlinearity remains optimal if the states are allowed to lie in a higher-dimensional space.  Furthermore, numerical evidence also suggests that for many other nonlinearities of the form \eq{nlin}, the optimal protocol for qubit states cannot be improved by embedding those states in a higher-dimensional space.  However, we do not have a proof that optimality can be achieved using only two dimensions.

A notable example where higher dimensions do offer an advantage is provided by the nonlinearity with $\nlin(x)=x^2-x^4$ (as considered in \cite{MW14}).  This nonlinearity has $\nlinred(z)=0$, so it cannot make two states of a qubit more distinguishable.  However, by embedding the states in three or more dimensions, it becomes possible to decrease their inner product.  In this case, numerical evidence suggests that there is no advantage to using more than four dimensions.

\section*{Acknowledgments}

AMC thanks Michael Foss-Feig, Alexey Gorshkov, and Eite Tiesinga for helpful discussions about the breakdown of the Gross-Pitaevskii approximation.
We also thank David Meyer and Thomas Wong for comments on a preliminary version of this paper.

This work was supported by the Canadian Institute for Advanced Research, the National Sciences and Engineering Research Council of Canada, the Ontario Ministry of Research and Innovation, and the US Army Research Office.


\end{document}